# Site dependent hydrogenation in Graphynes: A Fully Atomistic Molecular Dynamics Investigation

Pedro A. S. Autreto and Douglas S. Galvao
Instituto de Física 'Gleb Wataghin', Universidade Estadual de Campinas, 13083-970, Campinas, São Paulo, Brazil.

## ABSTRACT

Graphyne is a generic name for a carbon allotrope family of 2D structures, where acetylenic groups connect benzenoid rings, with the coexistence of sp and $sp^2$ hybridized carbon atoms. In this work we have investigated, through fully atomistic reactive molecular dynamics simulations, the dynamics and structural changes of the hydrogenation of α, β, and γ graphyne forms. Our results showed that the existence of different sites for hydrogen bonding, related to single and triple bonds, makes the process of incorporating hydrogen atoms into graphyne membranes much more complex than the graphene ones. Our results also show that hydrogenation reactions are strongly site dependent and that the sp-hybridized carbon atoms are the preferential sites to chemical attacks. In our cases, the effectiveness of the hydrogenation (estimated from the number of hydrogen atoms covalently bonded to carbon atoms) follows the α, β, γ-graphyne structure ordering.

## INTRODUCTION

The chemistry of carbon is very rich and this richness is due, mainly, to the fact that there are three possible different hybridizations (sp, $sp^2$ and $sp^3$) [1]. This characteristic allows a plethora of distinct allotropes, some of them discovered in the last few decades, such as fullerenes, nanotubes, and more recently, graphene [1,2].

Graphene is a two-dimensional array of hexagonal units $sp^2$ bonded C atoms, which has been studied theoretically since late 1940s as a model to describe some properties of graphite [3]. Although graphene exhibits several remarkable and unique mechanical and electronic properties, there are some difficulties to be overcome before a real graphene-based nanoelectronics becomes a reality. These difficulties are mainly related to its zero bandgap value, which precludes its use in some digital electronics applications, like transistors and diodes.

Diverse approaches have been tried to solve this issue, such as; application of strain, quantum confinement in nanoribbons, and chemical functionalization. Chemical functionalization is quite appealing since it can be used to open the electronic gap of structure as well as directly change the interaction of graphene with its environment. Other reasons that make functionalization attractive are that it can be also exploited to drug delivery, hydrogen storage, defect manipulation, magnetic devices, etc. [4,5]. However, controlled graphene functionalization has been only partially successful, with many issues still unsolved [6,7].

In part due to this, there is a renewed interest in other two-dimensional graphene structures, such as graphynes [8]. Graphyne (see Figure 1) is a generic name for a carbon allotrope family of 2D structures, where acetylenic groups connect benzenoid rings, with the coexistence of sp and $sp^2$ hybridized carbon atoms. Similarly to graphene, tubular structures can also exist [9-10]. Graphynes share some of the remarkable graphene properties, with the advantage that some forms are intrinsically non-zero bandgap structures [11]. Also, from the point of view of chemical functionalization, in contrast to graphene, graphynes exhibit non-equivalent functionalization sites.

In this work we have investigated, using fully reactive molecular dynamics, the structural and dynamical aspects of the hydrogenation processes of α (ALPHA), β (BETA), and γ (GAMMA) graphyne sheets (Figure 1).

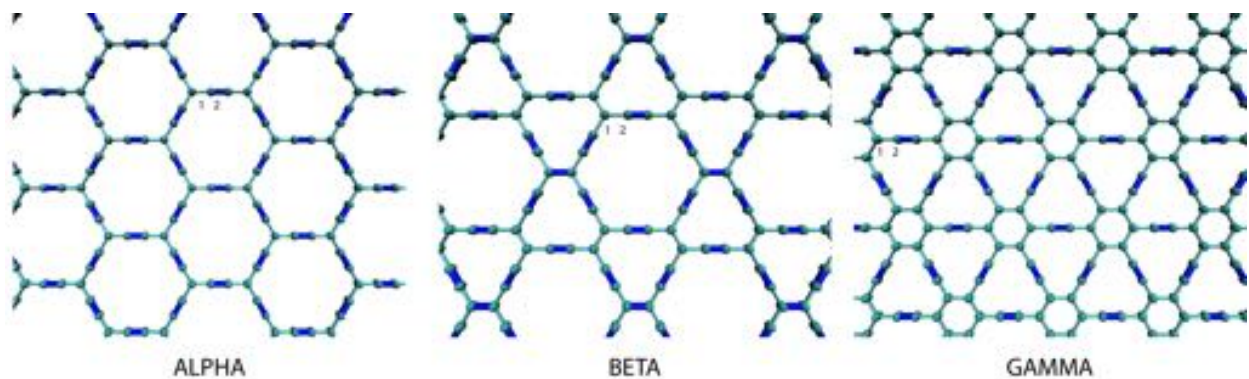

**Figure 1.** Optimized structures: (a) α-graphyne (ALPHA); (b) β-graphyne (BETA), and; (c) γ-graphyne (GAMMA). Green (light) and blue (dark) colors indicate single/double and triple bonds, respectively. Labeling 1 and 2 refer to carbon atoms in single/double and triple bonds, respectively.

**METODOLOGY**

We carried out fully atomistic reactive molecular dynamics (MD) simulations to investigate the structural and dynamical aspects of the hydrogenation of graphyne membranes. The simulations were carried out using reactive force fields (ReaxFF [12, 13]), as implemented in the Large-scale atomic/Molecular Massively Parallel Simulator (LAMMPS) code [14]. In order to speed up the simulations, we do not allow H-H recombination. We have also used the LAMMPS implemented Nosé-Hoover thermostat. The simulations were carried out at room temperature (300 K). Typical time for a complete simulation run was 4000 ps, with time-steps of 0.1fs.

The simulations were carried out considering the graphyne membranes immersed into an atmosphere of hydrogen atoms (Figure 2). Our systems consisted of isolated single-layer graphynes membranes (typical dimensions of 180 Å 180 Å (~ 6000 carbon atoms)), immersed into an atmosphere of atomic hydrogen atoms (~ 6000 H atoms).

ReaxFF is a well-tested reactive force field developed by van Duin, Goddard III and co-workers This force field allows simulations of many types of chemical reactions. Similarly to standard non-reactive force fields, like the popular MM3 [10-12], the system energy in ReaxFF, is also divided into different partial energy contributions associated with, amongst others, valence angle bending and bond stretching, and non-bonded contributions, such as, van der Waals and Coulomb interactions [10-12]. Also, it contains more terms, which can handle bond formation and dissociation (making/breaking bonds) as a function of bond order values. ReaxFF was parameterized against DFT calculations, being the average deviations between the heats of formation predicted by ReaxFF and the experiments equal to 2.8 and 2.9 kcal/mol, for non-conjugated and conjugated systems, respectively [10-12].

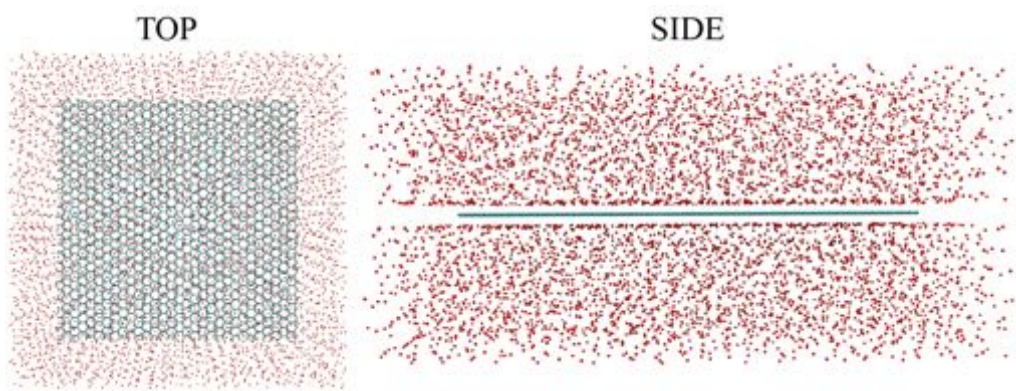

**Figure 2**. Simulation setup: Graphyne structures immersed into an atmosphere composed of hydrogen atoms. Top and side views, respectively.

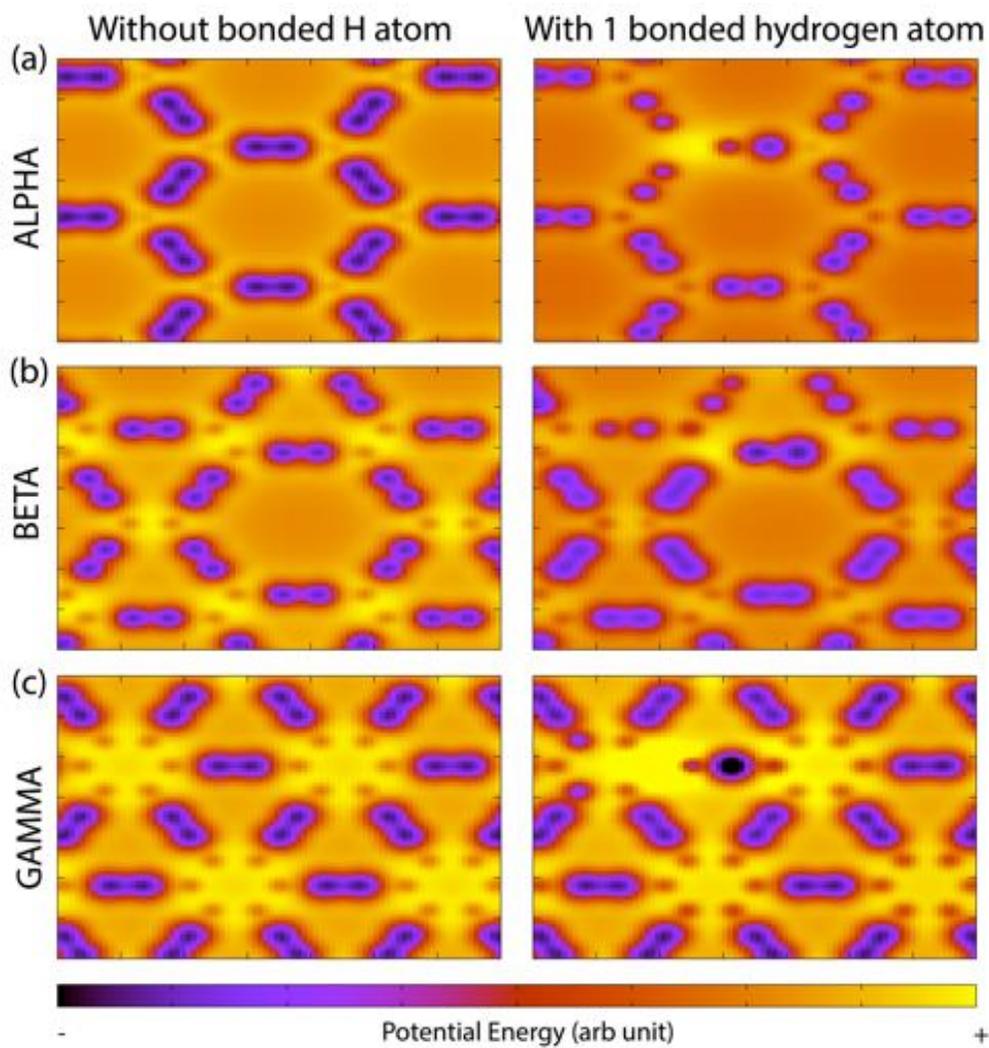

**Figure 3.** Potential maps for α, β, and γ-graphynes, and their hydrogenated forms. Results for an atom probe located at a height of 1.5 Å above the basal membrane plane.

**DISCUSSION**

In Figure 3 we present the calculated ReaxFF potential maps for the pristine and hydrogenated graphyne structures. These maps provide helpful and visual information about the potential experienced by a hydrogen atom when going to chemically interact with the membrane. The depth-well energy values are suggestive of which site would be first attacked by hydrogen atoms. From this Figure we can see that there are remarkable differences among the sites in the different graphynes membranes. The triple bonds are the most "attractive" places for hydrogenation, as we expected from a chemical point of view. We can also see that, although all structures present triple bonds, the level of reactivity is very different for the three structures. The hydrogenation effectiveness (estimated from the number of hydrogen atoms covalently bonded to carbon atoms), in principle, should follows the α, β, γ-graphyne structure ordering. After the first hydrogenation, these patterns are significantly altered.

The hydrogen effectiveness can be better estimated from Figure 4. As we can see from this Figure, the effectiveness follows the α, β, γ-graphyne structural ordering. α-graphyne has almost 60% of its Site 1-carbon type hydrogenated. Also, these reactivities are significantly site dependent: atoms with sp hybridization are the preferentially reaction sites for all graphynes forms. β is that one which presents the largest site differentiation (S2>>S1). γ-graphyne is the hardest form to be hydrogenated but the differences between S1 and S2 rates are similar to the ones observed for α-form, which is the easiest form to functionalize.

We have also observed that after a certain hydrogenation level, α-graphyne membranes started to structurally collapse from the shrinkage induced by bonding breaks (see Figure 5). The same occurs for β-graphyne, but a lower level, extensive part of the graphyne structure is preserved. Because of its resistance to be hydrogenated, γ-graphyne remains almost intact.

Another important result is that, in contrast to was reported to the case of graphene hydrogenation [6], we did not observe the formation of correlated domains (islands of hydrogenated carbons). This can be a consequence of the porous graphyne structure, which allows larger out-of-plane deformations (in comparison to graphene) and, consequently, an increase in the curvature, which results in an increased local chemical reactivity. Consistently, in the case of graphene fluorination (fluorine is more reactive than hydrogen atoms) the formation of these domains are also suppressed [7]. The results we obtained here for graphynes are process very similar to the ones recently reported to graphdiynes (graphyne-like structures with double acetylene linkages) [15].

**CONCLUSIONS**

In summary, we have investigated using fully atomistic reactive molecular dynamics simulations the structural and dynamical aspects of graphyne hydrogenation. Similar to graphene, graphynes are planar carbon allotropes where acetylenic groups connect benzenoid rings, with the coexistence of sp and $sp^2$ hybridized carbon atoms. In this work we have considered α, β, γ-graphyne forms.

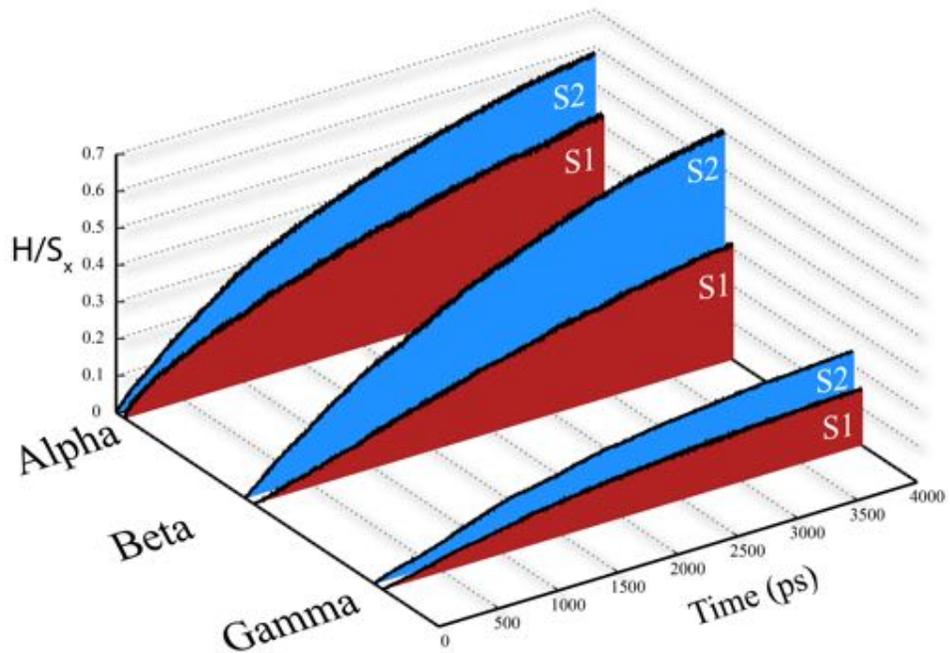

**Figure 4.** Number of incorporated H atoms as a function of time simulation. The S1 and S2 labels refer to atoms 1 and 2 shown in Figure 1, respectively.

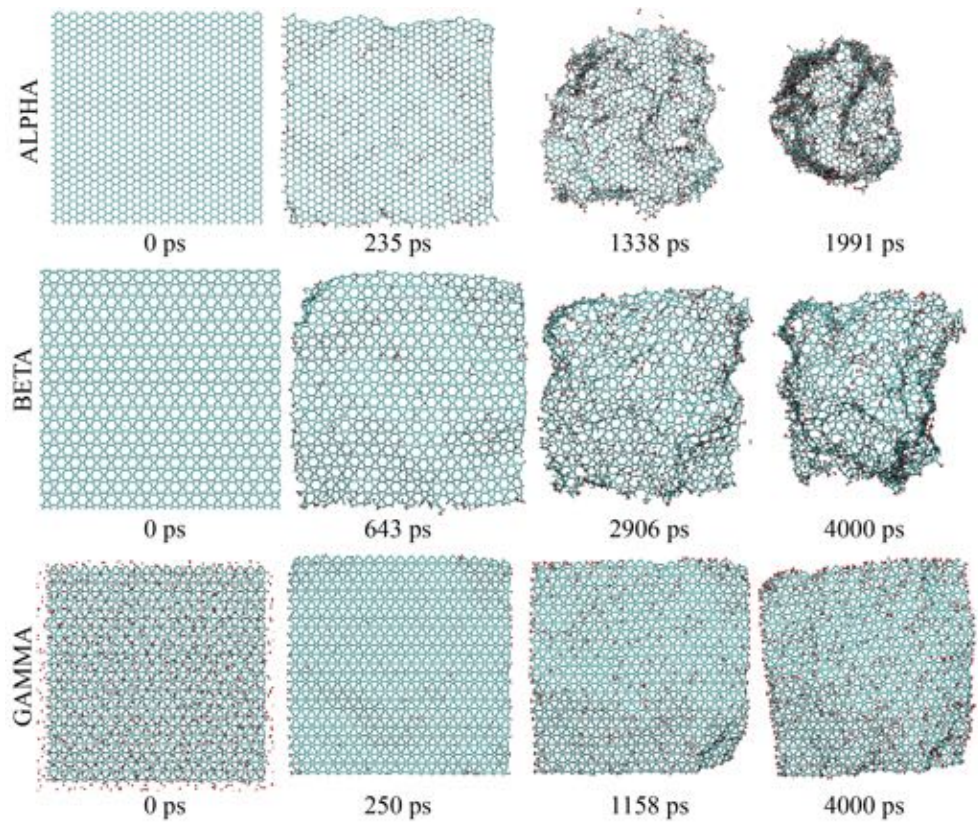

**Figure 5.** Snapshots of hydrogenation process of graphyne forms.

Our results showed that although all these structures contain triples bonds and share the planar arrangement, their hydrogenation dynamics is significantly different. For the investigated cases, the effectiveness of the hydrogenation (estimated from the number of hydrogen atoms covalently bonded to carbon atoms) follows the α, β, γ-graphyne ordering. However, the level of hydrogenation the structures can support before the destruction of the bi-dimensional topology follows an inverted ordering.

The existence of different sites for hydrogenation process, related to the co-existence of single, triple and resonant bonds, makes the process of incorporating hydrogen much more complex than the graphene ones. Another important result is that, in contrast to was reported to the case of graphene hydrogenation [6], we did not observe the formation of correlated domains (islands of hydrogenated carbons).

## ACKNOWLEDGEMENTS

This work was supported in part by the Brazilian Agencies CAPES, CNPq and FAPESP. The authors thank the Center for Computational Engineering and Sciences at Unicamp for financial support through the FAPESP/CEPID Grant # 2013/08293-7.